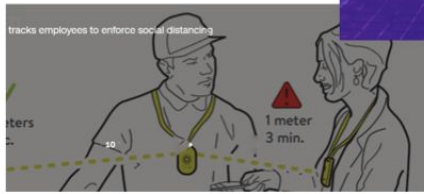

# Artificial Intelligence at the Edge
*A Computing Community Consortium (CCC) Quadrennial Paper*
*Elisa Bertino (Purdue University) & Sujata Banerjee (VMware Research)*

The Internet of Things (IoT) and edge computing applications aim to support a variety of societal needs, including the global pandemic situation that the entire world is currently experiencing and responses to natural disasters.

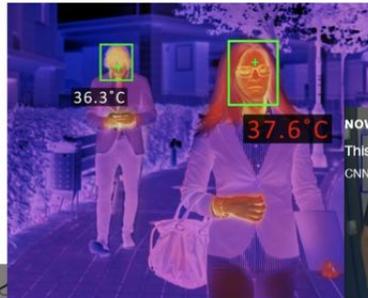
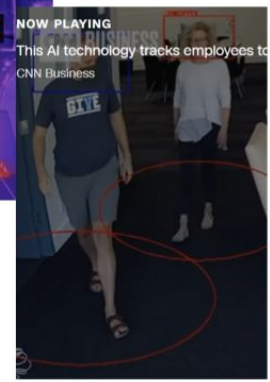

The need for real-time interactive applications such as immersive video conferencing, augmented/virtual reality, and autonomous vehicles, in education, healthcare, disaster recovery and other domains, has never been higher. At the same time, there have been recent technological breakthroughs in highly relevant fields such as artificial intelligence (AI)/machine learning (ML), advanced communication systems (5G and beyond), privacy-preserving computations, and hardware accelerators. 5G mobile communication networks increase communication capacity, reduce transmission latency and error, and save energy - capabilities that are essential for new applications. The envisioned future 6G technology will integrate many more technologies, including for example visible light communication, to support groundbreaking applications, such as holographic communications and high precision manufacturing. Many of these applications require computations and analytics close to application end-points: that is, at the edge of the network, rather than in a centralized cloud. AI techniques applied at the edge have tremendous potential both to power new applications and to need more efficient operation of edge infrastructure. However, it is critical to understand where to deploy AI systems within complex ecosystems consisting of advanced applications and the specific real-time requirements towards AI systems. There are many opportunities and associated challenges in edge AI as discussed below.

- **Analytics at the Edge:** Conventional approaches upload data to a cloud or other centralized servers; however such approaches may not be suitable due to the large volumes of data, response-time, privacy, and security requirements, especially when dealing with mobile systems and applications. Analytics at the edge may address such issues. Many privacy techniques for data analytics and machine learning have been proposed over the years. However, especially when addressing privacy issues, the best privacy techniques to use may depend on the specific analytics tasks to be executed and services to be provided, and on the specific situations (for example numbers of vehicles driving in a certain area). Therefore privacy-preserving analytics at the edge require adaptive frameworks able to dynamically select the most suitable privacy technique(s) based on the specific communication contexts and edge devices involved, the tasks to be carried out, and the specific tradeoff among degree of privacy, response time, and resource and energy costs. The same applies to analytics even when privacy is not a requirement, in that flexible analytic systems are needed to be able to dynamically adapt.
- **AI to Manage Infrastructure at the Edge:** AI will cover an increasingly important role in managing infrastructure - namely wireless communications and edge computing resources. Decisions concerning where to execute computational tasks at the edge will be complex as they have to take into account different factors, such as load on edge servers, signal strength for wireless communications, and mobility of users. For example, one may have to predict where a user will be moving in order to cache results at the proper edge server to minimize transmission costs and energy for the user devices. Past approaches based on operation research techniques are unable to deal with rapidly changing environments. Therefore AI approaches, like reinforcement learning, may be more suitable and need further investigation.
- **Real-time AI:** Many of the applications to be enabled by 5G and 6G have real-time requirements. Addressing such requirements is a complex problem dependent on the specific application, type(s) of device(s), type(s) of AI technique(s) to be used, and communication capabilities. Systems-of-systems approaches have to be designed for applications involving multiple edge devices - collecting different data and/or executing different applications and actions, and servers of different computing, communication, and mobility capabilities.
- **Resource-constrained AI:** Techniques have been proposed to reduce the size of AI models for deployment on small devices, such as IoT devices. Relevant techniques include model compression and pruning, but maintaining high accuracy of the models while applying these techniques is a challenge.
- **Distributed AI, Federated Learning:** For a variety of reasons outlined above, there is a need to train and build global models in a distributed manner without exchanging the raw data. Such federated learning approaches rely on secure private computations and reduce the communication overheads of transferring data over bandwidth constrained networks. At the same time, federated learning frameworks face many technical challenges stemming from heterogeneity in resource availability and data distributions.
- **Safe Edge AI:** Safety of AI techniques is paramount in many use cases, but becomes critical when the AI is used to perform actions that impact the physical space, as many IoT devices and robots have actuation capabilities. An example is a smart door lock that automatically unlocks a house door when the authorized user is at the door. Recent AI techniques allow devices to learn by themselves by exploring different actions and selecting the sequence of actions that maximize a user-specified reward, such as energy consumption. However such an approach may lead to unsafe/insecure states. It is therefore critical that explorations be constrained by safety/security policies; such constraining can follow different approaches, including modifying the reward functions to also consider safety/security and/or adding additional layers to the reinforcement learning system to filter out unsafe actions and add "safety" actions to the generated action sequences. A major challenge is the specification of proper safety/security policies, as they are context, application and user-dependent, especially when dealing with large numbers of autonomous collaborating devices. Transfer learning and adaptation techniques may need to be explored to address such a challenge.
- **Adaptive Context-sensitive AI:** When deploying AI at the edge, there are several alternatives concerning where to execute AI tasks, such as classifications and inferences. For example one can execute them at the device, at an edge server, or split the computations (see discussion above about real-time). The choice depends on a lot of parameters, like communication costs, signal strength (in the case of wireless networks), and energy consumption. So we need approaches in which these choices can be intelligently made by the edge system. Traditional optimization methods may not be suitable here and perhaps techniques like reinforcement learning could be an interesting approach.

### Recommendations

The opportunities presented by the use of AI at the edge are multi-faceted; benefitting from them will require a deeply interdisciplinary approach. Collaborative research efforts focused on specific application classes or domains should be encouraged, with an end-to-end systems view. The tensions and trade-offs between safety, security, privacy, performance, and cost need deep exploration; they cannot be explored in a piecemeal manner. While the ultimate goal should be to build generalizable solutions to the extent possible, as the field expands, it is imperative to demonstrate successful solutions, even in somewhat narrow contexts. Thus we advocate for funding at the federal level to seeding programs that explicitly encourage collaborative multi-disciplinary research between AI, systems, applications and human factors researchers with a focus on complete edge AI solutions.

### Reports:
- NSF report on Grand Challenges in Edge Computing, 2016.
- NSF/VMware Partnership on Edge Computing Data Infrastructure (ECDI) (nsf18540), 2018 The CCC Wide Area Data Analytics workshop talked about analytics at the edge: https://cra.org/ccc/wp-content/uploads/sites/2/2020/06/CCC-Wide-Area-Data-Analytics-Report.pdf

*This white paper is part of a series of papers compiled every four years by the CCC Council and members of the computing research community to inform policymakers, community members and the public on important research opportunities in areas of national priority. The topics chosen represent areas of pressing national need spanning various subdisciplines of the computing*

*research field. The white papers attempt to portray a comprehensive picture of the computing research field detailing potential research directions, challenges and recommendations.*

*This material is based upon work supported by the National Science Foundation under Grant No. 1734706. Any opinions, findings, and conclusions or recommendations expressed in this material are those of the authors and do not necessarily reflect the views of the National Science Foundation.*

*For citation use: Bertino E. & Banerjee S. (2020) Artificial Intelligence at the Edge. https://cra.org/ccc/resources/ccc-led-whitepapers/#2020-quadrennial-papers*